\documentclass[10pt,aps,prd,nofootinbib,superscriptaddress,twocolumn]{revtex4}
\usepackage{amssymb,amsmath,latexsym,graphics, graphicx,epsfig,multirow,comment,hyperref,appendix}

\newcommand{\gsim}{\lower.7ex\hbox{$\;\stackrel{\textstyle>}{\sim}\;$}}
\newcommand{\lsim}{\lower.7ex\hbox{$\;\stackrel{\textstyle<}{\sim}\;$}}

\def\OO{{\cal O}}

\newcommand{\keV}{\,\mathrm{keV}}

\newcommand{\bef}{\begin{figure}[htbp]\begin{center}}
\newcommand{\eef}{\end{center}\end{figure}}
\newcommand{\ISO}[2]{{}^{#2}\text{#1}}

\begin{document}

\pagestyle{plain}

\title{
\begin{flushright}
\mbox{\normalsize SLAC-PUB-14138}
\end{flushright}
\vskip 15 pt

The Poker Face of Inelastic Dark Matter: 
\\
Prospects at Upcoming Direct Detection Experiments}

\author{Daniele S. M. Alves}
\affiliation{
Theory Group, SLAC, Menlo Park, CA 94025}
\affiliation{
Physics Department, Stanford University,
Stanford, CA 94305}
\author{Mariangela Lisanti}
\affiliation{
Theory Group, SLAC, Menlo Park, CA 94025}
\affiliation{
Physics Department, Stanford University,
Stanford, CA 94305}
\author{Jay G. Wacker}
\affiliation{
Theory Group, SLAC, Menlo Park, CA 94025}

\begin{abstract}
The XENON100  and CRESST experiments will directly test the inelastic dark matter explanation for DAMA's 8.9$\sigma$ anomaly.  This article discusses how predictions for direct detection experiments depend on uncertainties in quenching factor measurements, the dark matter interaction with the Standard Model and the halo velocity distribution.  When these uncertainties are accounted for, an order of magnitude variation is found in the number of expected events at CRESST and XENON100.
%Given these unknowns, its is unlikely that the accumulated exposure from CRESST 2009 run will be able to exclude the iDM hypothesis. 
%In contrast, XENON100 will resolve the DAMA/iDM debate, and in case of a positive signal, CRESST and XENON100 combined data might shed light on many dark matter particle and astrophysical properties.   
\end{abstract}
 \maketitle

Predictions for direct detection experiments require a wide-range of assumptions concerning the astrophysical properties of the dark matter, as well as its interactions with the Standard Model (SM).  These theoretical uncertainties are compounded by additional experimental challenges that arise from the nature of low energy experiments.  
Ultimately, it is necessary to know the  scattering rate for dark matter off SM nuclei in detectors. 
There are many unknown physical quantities that go into this prediction and they are often benchmarked to values in specific studies.  However, in light of a potential signal, the verification process requires a more systematic study of these unknowns in order to have a complete picture of the range of consistent theories.

Inelastic Dark Matter (iDM) serves as a case study for this new treatment of uncertainties and  shows how marginalizing over astro and particle physics quantities leads to at least an order of magnitude variation in detection prospects at upcoming experiments.  Inelastic dark matter is an elegant explanation for DAMA's on-going 8.9$\sigma$ annual modulation signal \cite{DAMA}, resolving the inconsistency of this signal with the plethora of null direct detection experiments \cite{CDMS, ZEPLINII, ZEPLINIIIiDM, CRESSTII, XENON10iDM, EDELWEISS}.  In the iDM framework, a dark sector particle up-scatters off the detector's target nucleus to a higher mass state \cite{TuckerSmith:2001hy}.  To explain the DAMA anomaly, an  $\OO(100 \text{ keV})$ mass splitting is required for weak-scale dark matter.  

IDM requires a minimum velocity to up-scatter to the more massive state, which 
depends on the mass of the target nucleus, $m_\text{N}$, the reduced mass of the nucleus-dark matter system, $\mu_N$, the mass splitting, $\delta$, and the recoil energy, $E_R$, of the nucleus:
%5
\begin{equation}
v_{\text{min}} = \frac{1}{\sqrt{2 m_\text{N} E_R}} \Big( \frac{m_\text{N} E_R}{\mu_N} + \delta\Big).
\end{equation}
The detection rate \cite{Lewin:1995rx} depends on $v_{\text{min}}$ through 
\begin{equation}
\frac{dR}{dE_R} =\frac{\rho_0}{m_{\text{dm}}m_N} \int_{v_{\text{min}}}^{v_{\text{esc}}}  f(\vec{v} + \vec{v}_E(t)) v \frac{d \sigma}{d E_R} d^3v, 
\end{equation}
where $f(\vec{v})$ is the local velocity distribution function (vdf) for the dark matter halo in the galactic frame, and $\vec{v}_E(t)$ accounts for the boost to Earth's rest frame \cite{Schoenrich:2009bx}.  The differential scattering rate is larger for heavier target nuclei because $v_{\text{min}}$ is reduced.  

The spin-independent differential cross section can be parameterized as 
\begin{eqnarray}
\frac{d\sigma}{dE_R} =\frac{m_N\sigma_\text{n}}{2\mu_{\text{n}}^2 v^2} (f_\text{p} Z+f_\text{n}(A-Z))^2  |F_{\text{dm}}(q^2) F_N(q^2)|^2,
\end{eqnarray}
where $\sigma_\text{n}$ is the dark matter-nucleon cross section at zero momentum transfer, $\mu_{\text{n}}$ is the dark matter-nucleon reduced mass, and $q^2 = 2 m_N E_R$ is the momentum transfer.   The constants $f_{\text{p},\text{n}}$ parameterize the coupling to the proton and neutron, respectively, and are set to $f_\text{p} = f_\text{n} = 1$ throughout. The dependence of the cross section on the nuclear recoil energy comes from the dark matter and nuclear form factors, $F_{\text{dm}}(q^2)$ and $F_N(q^2)$.  $F_{\text{dm}}(q^2)$ describes non-trivial behavior at low momentum transfer in models where higher dimensional operators contribute to the scattering  \cite{Feldstein:2009tr,Chang:2009yt, Lisanti:2009am}.  
$F_\text{N}(q^2)$ is the Helm/Lewin-Smith nuclear form factor \cite{Lewin:1995rx}.  Analytic approximations to the nuclear form factor can have substantial errors, particularly for heavy nuclei such as $\ISO{W}{184}$.  The Helm/Lewin-Smith form factor is better behaved than other Helm parameterizations, but can still give errors of 25\% for $\ISO{W}{184}$ in the range $E_R = 10-40$ keVnr \cite{Duda:2006uk}. Around 100 keVnr, these errors can be as large as 60\%.  The impact of nuclear form factor uncertainties on predictions for direct detection has been addressed in the literature \cite{SchmidtHoberg:2009}.  This work explores other sources of uncertainty, and for the rest of this paper the Helm/Lewin-Smith form factor is adopted.

Typically, the vdf is taken to be Gaussian, isothermal and isotropic in the galactic frame.  This `Standard Halo Model' (SHM) is parameterized as
\begin{equation}
\label{SHM}
f(v) \propto \Big( e^{-v^2/v_0^2} - e^{-v^2_{\text{esc}}/v_0^2}\Big) \Theta(v_{\text{esc}} - v),
\end{equation}
where $v_{\text{esc}}$ is the galactic escape velocity and $v_0$ is the velocity dispersion. The range of escape velocities is constrained by the RAVE stellar survey: $480 \leq v_{\text{esc}} \leq 650$ km/s \cite{Smith:2006ym}, and no constraints are placed on $v_0$.  The standard procedure when evaluating direct detection rates is to assume a SHM distribution with benchmarked values for $v_0$ and $v_{\text{esc}}$.  The solid blue curve in Fig.~\ref{fig:recoilspectrum} shows the expected tungsten recoil spectrum for this vdf; the bulk of events occur between 10-40 keVnr.  

While previous studies have looked at the effect of varying $v_0$ and $v_{\text{esc}}$ within the SHM \cite{McCabe:2010zh,Kopp:2009qt}, none have fully marginalized over both dark matter and halo profile uncertainties.  In addition, numerical N-body simulations indicate significant departure of the vdf from the SHM hypothesis \cite{Diemand:2006ik, MarchRussell:2008dy, Fairbairn:2008gz, Kuhlen:2009vh}, especially in the high velocity tail.  Because very little is known about either the vdf or the dark matter model, experimental analyses should be designed to cover a wide range of possibilities.  In this paper, a scan over the parameter space for iDM is performed, where we marginalize over the dark matter parameters $(m, \delta, \sigma)$ and halo velocity parameters $(v_0, v_{\text{esc}}, \vec{v}_\text{stream})$, and set constraints by a global $\chi^2$ analysis  \cite{Lisanti:2009vy}.\footnote{In some circumstances, maximum gap techniques provide tighter limits than Poisson statistics for null experiments \cite{Yellin:2002xd, SchmidtHoberg:2009}.  However, Poisson statistics are used in this paper due to the complexity of combining a $\chi^2$ for DAMA with multiple max-gap tests to get a global limit.}

The predicted number of events at CDMS \cite{CDMS}, ZEPLIN-II \cite{ZEPLINII}, ZEPLIN-III \cite{ZEPLINIIIiDM}, CRESST-II \cite{CRESSTII}, XENON10 \cite{XENON10iDM},  EDELWEISS \cite{EDELWEISS},  and the XENON100 calibration run \cite{XENON100} are included in the $\chi^2$ as well as  the annual modulation amplitude in the first twelve bins of DAMA (2-8 keVee) \cite{DAMA}.  The high energy bins from 8-14 keVee are combined into a single bin with modulation amplitude -0.0002$\pm$0.0014 cpd/kg/keVee.  Any model that  over-predicts the number of events at the null experiments by $2\sigma$ is excluded.

\section*{Inelastic Dark Matter at CRESST}

The standard assumption is that the iDM interpretation of DAMA's signal can be confirmed or refuted by any experiment with a target nucleus heavy enough to cause the inelastic transition \cite{Chang:2008gd}.  For DAMA, the inelastic transition occurs through scattering off the $\ISO{I}{127}$ nucleus in the NaI(Tl) target. Naively, any experiment with a target mass greater than $\ISO{I}{127}$ could provide a sufficient test.  Two upcoming experiments fall into this category: XENON100 \cite{XENON100} and CRESST \cite{CRESST09, WONDER}, which use $\ISO{Xe}{131}$ and $\ISO{W}{184}$, respectively.  XENON100 has currently released results from a calibration run of 11.2 live days during Oct-Nov 2009.  They report 161 kg-d effective exposure and have observed no events in their acceptance region between 4.5 - 40 keVnr.  CRESST, which consists of nine detectors of CaWO$_4$ and one detector of ZnWO$_4$, has shown preliminary results in the energy window from 10 - 40 keVnr obtained from summer 2009 until the present; however, the exposure was not reported \cite{WONDER}.

CRESST provides a unique experimental environment for testing iDM because it has the heaviest target nucleus of all current direct detection experiments.  $\ISO{W}{184}$ is expected to be highly sensitive to inelastic scattering because its velocity threshold is a factor $\sqrt{m_{\text{I}}/m_{\text{W}}}\sim0.83$ lower than iodine.  As a result, a larger fraction of the halo can up-scatter off of $\ISO{W}{184}$ and one would expect a larger scattering rate compared to lighter targets.  However, an additional complication arises due to the large radius $R_{\text{W}}$ of a  $\ISO{W}{184}$ nucleus.  In particular, when the momentum transfer is $q \sim 1/R_{\text{W}}$, the dark matter probes the size of the nucleus and the scattering is no longer coherent.  Therefore, the scattering rate is suppressed at recoil energies $E_R \propto 1/(2 m_\text{N} R_{\text{W}}^2)$.  This suppression occurs at lower recoil energies for $\ISO{W}{184}$, as compared to $\ISO{Xe}{131}$ (55 keVnr versus 90 keVnr, respectively).

\begin{figure}[tb] %  figure placement: here, top, bottom, or page
   \centering
  \includegraphics[width=3.25 in]{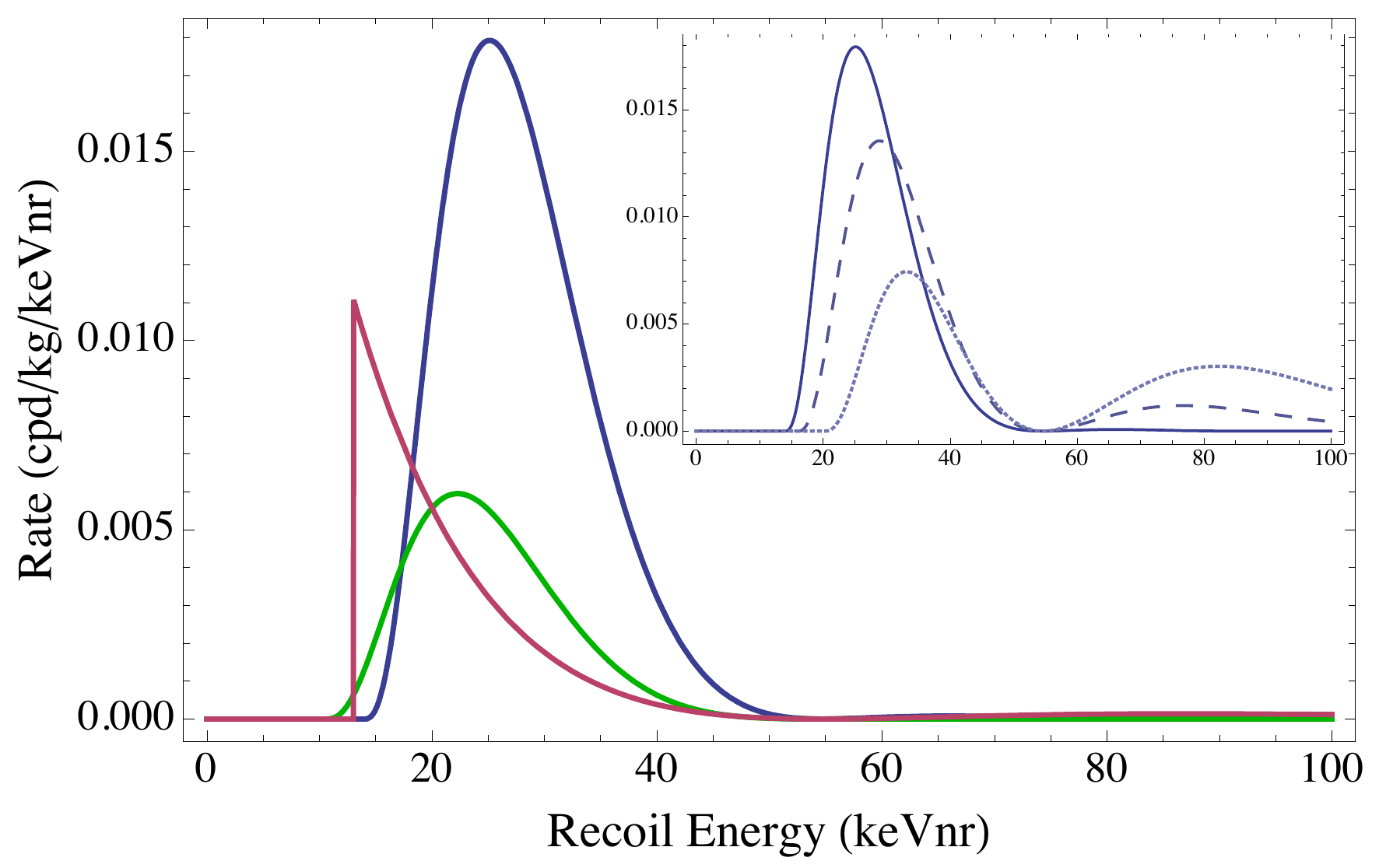}
   \caption{CRESST spectra for: regular iDM (blue), FFiDM with $F_\text{dm}\propto E_R$ (green), DM stream (red) for $Q_{\text{I}}=0.085$. Inset: $Q_{\text{I}}=0.07$ (dashed), $Q_{\text{I}}=0.06$ (dotted). }
   \label{fig:recoilspectrum}
\end{figure}

The fact that the first zero of the $\ISO{W}{184}$ form factor occurs at such a low recoil energy highlights an important challenge for CRESST.  The typical recoil energy an iDM particle deposits on a $\ISO{W}{184}$ nucleus is $\OO(\delta\hspace{0.02in}\mu/m_{\text{W}})\sim\OO(50~\text{keVnr})$, where the signal is suppressed due to loss of coherence.  The form factor suppression is evident in the recoil spectra of Fig.~\ref{fig:recoilspectrum}, which also illustrates the effects of three additional sources of uncertainties:  DAMA's energy calibration, the dark matter interaction with the SM, and the velocity distribution profile.  Variations in any of these three sources can significantly alter iDM's rate at CRESST and ultimately affect the final exposure that will be required for CRESST to exclude the DAMA iDM hypothesis.

\subsubsection*{DAMA's Energy Calibration}
 
DAMA only detects its nuclear recoil events with scintillation light. Nuclear recoils typically deposit only a small fraction of their energy into scintillation. The quenching factor for $\ISO{I}{127}$ , $Q_{\text{I}}$, relates the measured electron equivalent energy (given in keVee) to the nuclear recoil energy (given in keVnr):
\begin{equation}
E_{\text{ee}}= Q_{\text{I}}(E_{\text{nr}})\;E_{\text{nr}},
\end{equation}
where the energy dependence of the quenching factor is left explicit.  Most studies assume a constant quenching factor for iodine from $\sim 10- 100$ keVnr, with the standard value taken to be $Q_{\text{I}}=0.085$. However, there are large experimental uncertainties in measurements of $Q_{\text{I}}$ \cite{Bernabei:2003za}.  The four primary ones \cite{Bernabei:1996vj,Pecourt:1998jg,Tovey:1998ex,Fushimi:1993nq} give $0.05\le Q_{\text{I}} \le 0.10$.
%%
%\begin{center}
%\begin{tabular}{ | c | c | c| }
%\hline
%Recoil Energy (keVnr) & $Q_{\text{I}}$ & Reference \\ \hline
%22-330 & 0.09$\pm$ 0.01& \cite{Bernabei:1996vj} \\ \hline
%40-100 & 0.08$\pm$ 0.02 &\cite{Pecourt:1998jg}\\ \hline
%10-71 & 0.086$\pm$ 0.007 &\cite{Tovey:1998ex}\\ \hline
%40-300 & 0.05$\pm$ 0.02 &\cite{Fushimi:1993nq}\\ \hline
%\end{tabular}
%\end{center}
%%
The study in \cite{Tovey:1998ex} gives the smallest error, however its measurements are calibrated with 60 keV gamma rays, in contrast to the 3.2 keV electrons that DAMA uses  \cite{Bernabei:1996vj}.  This difference reduces the central value of \cite{Tovey:1998ex} by roughly 10\% and induces larger systematic effects.  

Lowering iodine's quenching factor effectively shifts DAMA's signal to higher nuclear recoil energies, favoring slightly larger values for the iDM mass splitting ($100 \lesssim \delta \lesssim 180$ keVnr for $Q_{\text{I}}=0.06$). Consequently, the predicted signal at other experiments is also shifted to higher nuclear recoil energies. In addition, the spectral shape is broadened, because DAMA's reported rate is in units of cpd/kg/keVee. 

\begin{figure}[tb] %  figure placement: here, top, bottom, or page
   \centering
   \includegraphics[width=3.25in]{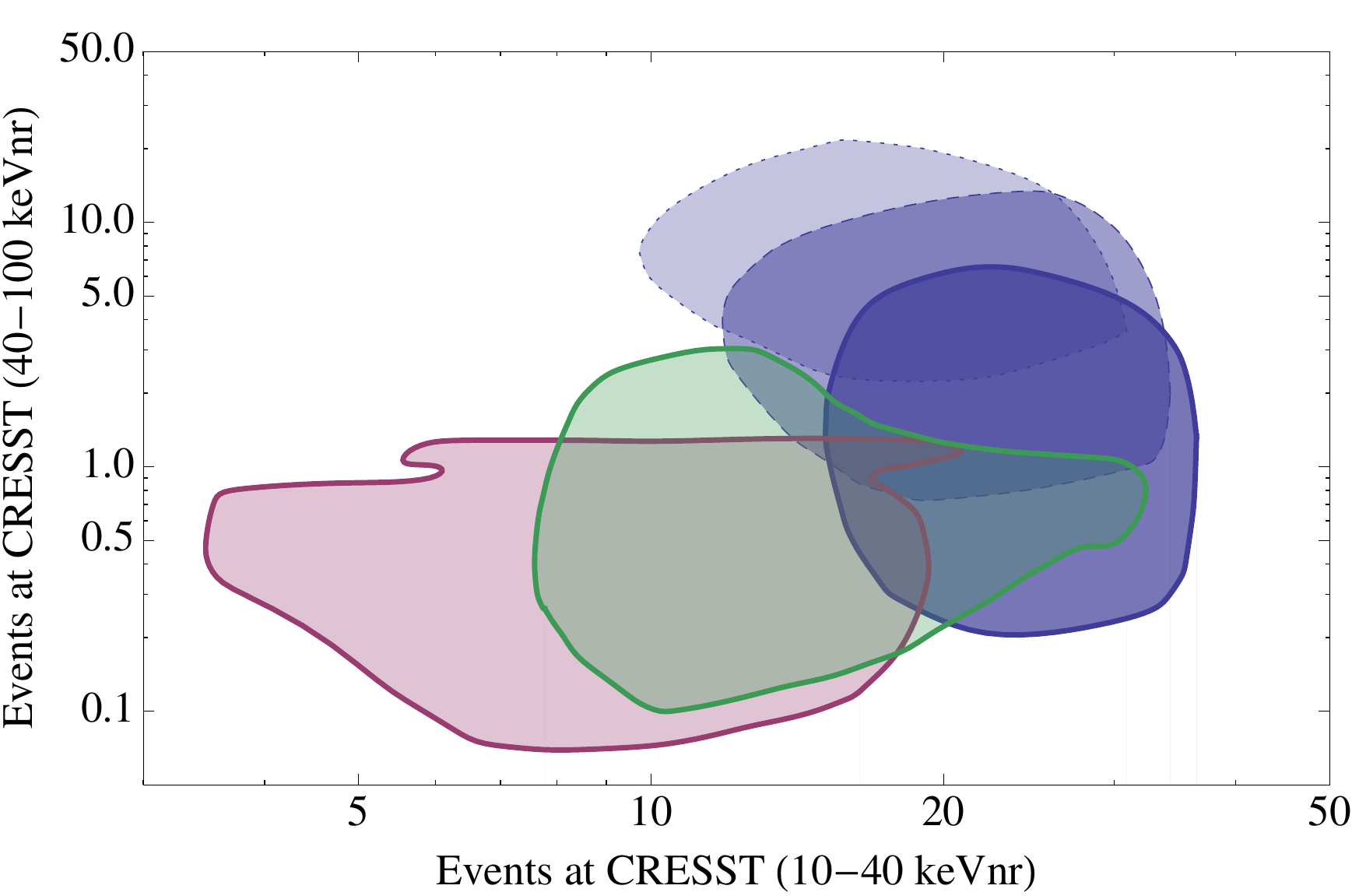} 
   \caption{Average counts at CRESST per 100 kg-d for regular iDM (blue), FFiDM with $F_\text{dm}(q)\propto E_R$ (green), and DM streams (red). The effect of lowering the quenching factor is illustrated for $Q_{\text{I}} = 0.07$ (dashed blue) and $ Q_{\text{I}}=0.06$ (dotted blue). The contours enclose all points with $\chi^2 \leq 18$.}
   \label{Fig: Quench}
\end{figure}

Fig.~\ref{fig:recoilspectrum} illustrates how the $\ISO{W}{184}$ recoil spectrum changes as $Q_{\text{I}}$ is reduced from 0.085 to 0.06, and Fig.~\ref{Fig: Quench} shows the average annual rate for CRESST's low and high energy range, assuming a SHM profile.  The shift of the iDM signal to higher recoils translates into a significant reduction in CRESST's average annual rate in the low recoil window of $10 - 40$ keVnr and a substantial enhancement in its rate in the high recoil range from $40-100$ keVnr.  

The effect of marginalizing over both the particle and halo parameters is substantial.  As an illustration, if only the mass and cross section are marginalized over (with fixed $\delta = 120$ keV, $v_0 = 220$ km/s,  $v_{\text{esc}} = 550$ km/s), then the average counts at CRESST per 100 kg-d range from $\sim 17 - 36$ in the lower energy window and $\sim 0.3 -0.8$ in the high energy window for $Q_I = 0.085$.  At XENON100, the predicted number of events (per 1000 kg-d)  ranges from $\sim 33 - 100$, covering only a subset of all allowed possibilities illustrated in Fig.~\ref{fig:SHMscan}.

\subsubsection*{Dark Matter Interaction}

The identity of iDM is unknown and its interactions with the SM may not occur through renormalizable operators. Non-renormalizable operators typically result in matrix elements with non-trivial dependence on the momentum transfer $q$.  These can be parameterized by an effective DM form factor \cite{Pospelov:2000bq, Sigurdson:2004zp,Dobrescu:2006au,Duda:2006uk,Feldstein:2009tr,Chang:2009yt,Kopp:2009qt},
\begin{equation}
F_{\text{dm}}(q) =\sum\limits_{n,m}c_{n,m}\frac{(q^0)^n|\vec{q}\hspace{0.025in}|^m }{\Lambda^{n+m}} +\ldots
\end{equation}
where $q^0=E_R$, $|\vec{q}\hspace{0.03in}|=\sqrt{2m_NE_R}$, and $\Lambda$ is an arbitrary mass scale.
Standard iDM assumes that the constant $n,m=0,0$ term dominates the expansion. Models that have an interaction mediator with mass lighter than $\OO(|\vec{q}\,|)$ are dominated by $c_{0,-2}$.  Composite iDM models  have $c_{0,1}\ne 0$ \cite{Alves:2009nf,  Kaplan:2009de,Kribs:2009fy}.  
Form factors that are dominantly $n\ne 0$ can be realized through dipole or other tensor interactions \cite{Sigurdson:2004zp,Dobrescu:2006au}.

Standard iDM ({\it i.e.}, $n,m=0,0$) and models with $n=0, m\ne 0$ have comparable rates at CRESST because the ratio of predicted events between these two scenarios  scales as $N_{0,m}/N_{0,0} \simeq (m_\text{W} E_{\text{W peak}}/m_{\text{I}} E_{\text{I peak}})^{2m} \simeq 1$.   DAMA's spectrum peaks at $E_{\text{I peak}}\simeq 35\keV$ while the tungsten spectrum at CRESST peaks at  
$E_{\text{W peak}}\simeq 25\keV$.
In contrast,  interactions with $n\ne 0, m=0$ predict substantially smaller rates at CRESST:  $N_{n,0}/N_{0,0} \simeq ( E_{\text{W peak}}/E_{\text{I peak}})^{2n} \simeq (0.5)^{n}$. This effect is illustrated in Fig.~\ref{Fig: Quench} for $n,m=1,0$.

\subsubsection*{Dark Matter Velocity Distribution}

There is little direct observational evidence for the DM density profile, and the velocity distribution is highly uncertain.  While most studies assume a Maxwell-Boltzmann vdf  (\ref{SHM}), N-body simulations indicate that this ansatz does not adequately parameterize the vdf  \cite{Kuhlen:2009vh}.  The iDM spectrum is particularly sensitive to changes in the tail of the velocity distribution profile, which can arise from velocity anisotropies or from dark matter substructure that has recently fallen into the galaxy \cite{Diemand:2006ik, Lang:2010cd}.  A vdf in which the high velocity tail  is dominated by a stream of dark matter illustrates how changes in the local vdf alter iDM predictions.  This scenario can significantly lower the number of expected events; other possibilities for the velocity profile will result in numbers of events between the SHM and stream expectations.    

Streams of dark matter are characterized by low velocity dispersion \cite{Diemand:2006ik}.  Here, streams will be parameterized as  dispersionless vdfs that have an arbitrary incident angle.  The distribution profile is $f(\vec{v}) = \delta^3(\vec{v} - \vec{v}_{\text{stream}})$, with $\vec{v}$ and $\vec{v}_{\text{stream}}$ given in the frame of the sun.  The differential scattering rate is obtained after boosting to the Earth's frame, and depends on the recoil energy through
\begin{equation}
\frac{dR}{dE_R} \propto \frac{\Theta(| \vec{v}_{\text{stream}}-\vec{v}_E(t)| - v_{\text{min}})}{| \vec{v}_{\text{stream}} - \vec{v}_E(t)|} |F_{\text{N}}(q^2)|^2,
\end{equation}
where $\vec{v}_E(t)$ is the Earth's velocity in the frame of the solar system.  The rate would be constant if not for the nuclear form factor that shapes the distribution and yields a highly peaked spectrum as illustrated for tungsten in Fig.~\ref{fig:recoilspectrum}.

The stream's velocity and incident angle relative to the Earth are marginalized over and  constraints on the phase and higher harmonics of the annual modulated rate are applied.  These are set by the spectral decomposition of DAMA's modulated rate, which restricts the modulation to peak at May $24^\text{th}$ $\pm~7.5$ days and constrains the power spectrum at a frequency $\omega=2\text{ yr}^{-1}$ to be $P(2\text{ yr}^{-1})\lsim0.05P(\text{yr}^{-1})$ \cite{DAMA}.  We apply these additional cuts on the models that survive the $\chi^2$ minimization.

\begin{figure}[tb] %  figure placement: here, top, bottom, or page
   \centering
   \includegraphics[width=3.25in]{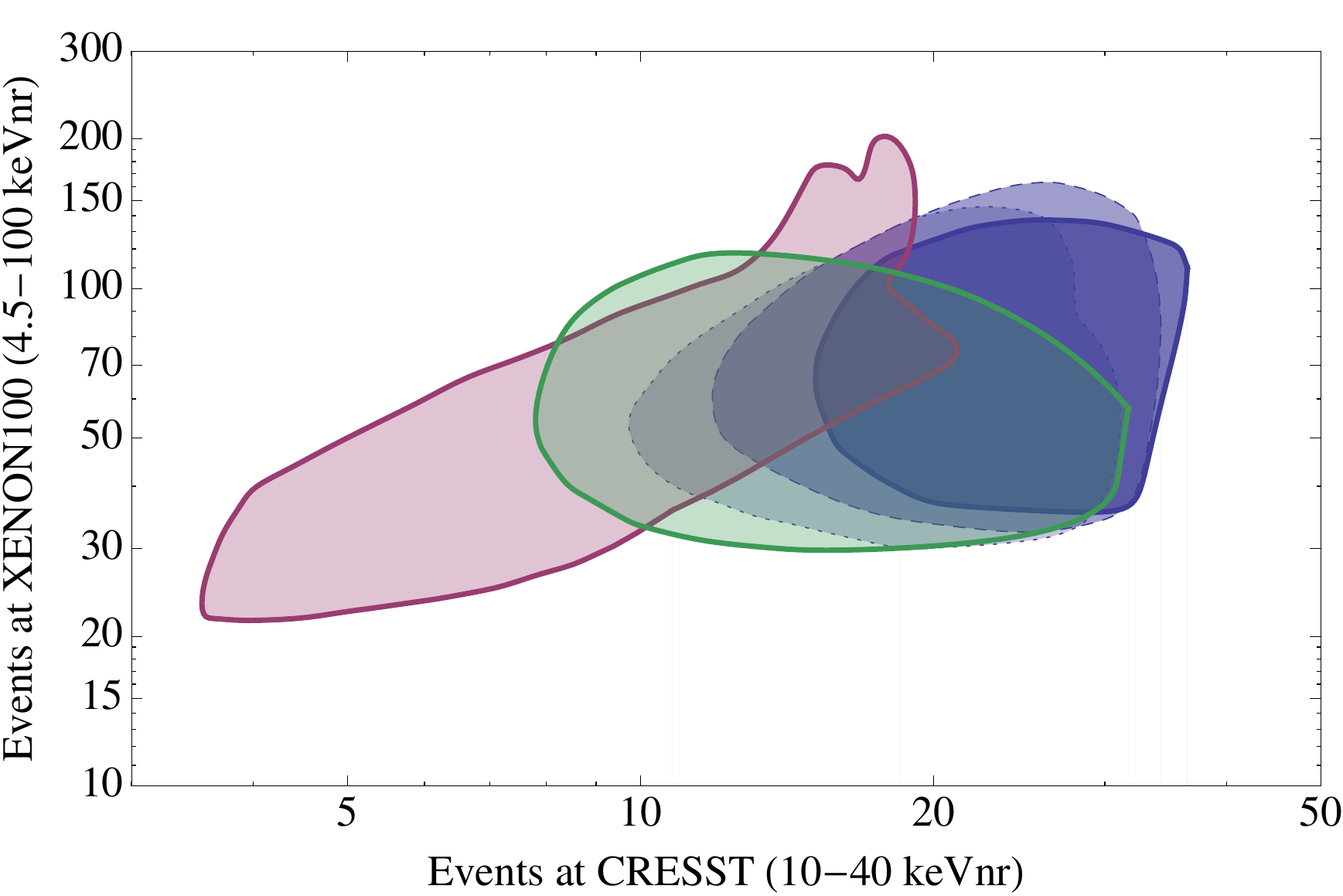} 
   \caption{ Average counts at CRESST (per 100 kg-d) versus XENON100 (per 1000 kg-d) for regular iDM (blue), FFiDM with $F_\text{dm}(q)\propto E_R$ (green), and DM streams (red). The effect of lowering the quenching factor is illustrated for $Q_{\text{I}} = 0.07$ (dashed blue) and $ Q_{\text{I}}=0.06$ (dotted blue). The contours enclose all points with $\chi^2 \leq 18$.}
   \label{fig:SHMscan}
\end{figure}

Fig.~\ref{Fig: Quench} shows the iDM predictions for CRESST when the tail of the vdf is dominated by a dispersionless DM stream.  When the dark matter stream is nearly head-on in the summer, very few events are expected in the fall and winter months, which is consistent with the late-year running of the XENON100 calibration run.  Marginalizing over stream parameters shows that velocities $v_\text{stream}\sim$ 400 km/s are favored, and that the DM incident angle with respect to the velocity of the Earth on June 2 can be as large as $75^\circ$, although 90\% of the consistent models have incident angle $\theta_\text{in}<50^\circ$ and 75\%  have $\theta_\text{in}<36^\circ$.  Fig.~\ref{Fig: Quench} shows that
the annual average rate at CRESST for dark matter streams can deviate dramatically from the SHM case and highlights the importance of marginalizing over parameters and considering different velocity profiles when making predictions for direct detection experiments.  

\section*{XENON100 Prospects}

The previous section discussed how uncertainties in energy calibration, dark matter interactions, and vdfs significantly affect the range of predicted events at CRESST.  For $\OO(100$ kg-d) exposures, the number of events in the low energy window can vary over an order of magnitude from $3 -30$, while the number in the high energy window ranges from $ 0.1 -10$.  It is evident that having significant exposure over the full nuclear recoil band where the iDM signal is expected to dominate (10-100 keVnr)  is essential for refuting or confirming a potential signal.  

The XENON100 experiment will accumulate large exposures $\OO(3000$ kg-d) in their current data run.  Compared to CRESST, it has better coverage of the relevant nuclear recoil band because the  $\ISO{Xe}{131}$ target is not  affected by form factor suppression at energies below $90$ keVnr.  
The dominant variation in predictions for  spin-independent iDM scattering rates in $\ISO{Xe}{131}$ arises because DAMA's modulation fraction is not known and only constrained to be greater than 2\% \cite{DAMA}.   Larger modulation fractions at DAMA imply a proportionately smaller rate at XENON100.
 
In their calibration run,  XENON100 demonstrated the potential to be a ``zero background" experiment.  The fact that no events were seen in their acceptance region implies a lower bound on the modulation fraction for iDM of $\OO(40\%)$.  Their next data release, which will include summer data, will directly test the  iDM parameter space still allowed.  Fig.~\ref{fig:SHMscan} illustrates the average number of expected events for XENON100 (per 1000 kg-d), including the uncertainties in $Q_{\text{I}}$, the DM form factor, and the vdf.  Again, there is an order of magnitude uncertainty, with the predictions ranging from 20-200 counts per 1000 kg-d.  For the expected exposure of XENON100's data release, the minimum number of events predicted by iDM is $\sim60$, which will be enough to confirm or refute the spin-independent iDM scenario in a conclusive and model-independent manner.  

\section*{Conclusions}

The process of testing the DAMA anomaly highlights many of the challenges inherent to direct detection experiments.  In addition to determining the properties of the unknown dark matter particle, direct detection experiments must also consider the unknown flux of the incident dark matter, as well as uncertainties in converting a signal from one target nucleus to another.

The predictions for both the CRESST 2009 run and XENON100 2010 run show an order of magnitude uncertainty. 
The nuclear form factor for $\ISO{W}{184}$, when combined with additional theoretical and experimental uncertainties, will likely prevent CRESST from refuting the iDM hypothesis with an exposure of $\OO(100~\text{kg-d})$  in a model-independent manner.  XENON100, on the other hand, will be able to make a definitive statement about  a spin-independent, inelastically scattering dark matter candidate.  Still, the CRESST 2009 data can potentially confirm iDM for a large range of parameter space. In case of a positive signal,  the combined data from CRESST and XENON100 will start probing the properties of the Milky Way DM profile and the interaction of the SM with the dark matter.

\section*{Acknowledgements}

We thank Neal Weiner for useful discussions, especially on quenching factors.  We would also like to thank Fabio Capella for clarification on the DAMA power spectrum and Rafael Lang for helpful explanations about backgrounds in CRESST's high energy regime.  DA, ML and JGW are supported by the US DOE under contract number DE-AC02-76SF00515 and receive partial support from the Stanford Institute for Theoretical Physics.  ML is supported by an NSF fellowship.  JGW is partially supported by the US DOE's Outstanding Junior Investigator Award.  As we neared completion of this paper, similar ideas were discussed at \cite{WeinerTalk}.

%%%%%%%%%%%%%%%%%%%%%%%%%%%%%

\end{document}